\documentclass[runningheads]{llncs}
\usepackage[T1]{fontenc}
\usepackage{graphicx}
\usepackage{multirow, xcolor, colortbl}
\usepackage{amssymb, amsmath, bbding}
\newcommand{\secondbest}[1]{\underline{#1}} 
\newcommand{\best}[1]{\textbf{#1}} 
\usepackage{cite}
\begin{document}
\title{MDPG: Multi-domain Diffusion Prior Guidance for MRI Reconstruction}
%
%
%


\author{Lingtong Zhang, Mengdie Song, Xiaohan Hao, Huayu Mai, Bensheng Qiu\inst{(}\Envelope\inst{)}}
\authorrunning{L. Zhang et al.}
\institute{School of Information Science and Technology,\\ University of Science and Technology of China, Hefei, China \\
    \email{\{zhanglingtong,smengdie,hxh045,mai556\}@mail.ustc.edu.cn} \\
    \email{bqiu@ustc.edu.cn}}

\maketitle              
\begin{abstract}
Magnetic Resonance Imaging (MRI) reconstruction is essential in medical diagnostics. As the latest generative models, diffusion models (DMs) have struggled to produce high-fidelity images due to their stochastic nature in image domains. Latent diffusion models (LDMs) yield both compact and detailed prior knowledge in latent domains, which could effectively guide the model towards more effective learning of the original data distribution. Inspired by this, we propose Multi-domain Diffusion Prior Guidance (MDPG) provided by pre-trained LDMs to enhance data consistency in MRI reconstruction tasks. Specifically, we first construct a Visual-Mamba-based backbone, which enables efficient encoding and reconstruction of under-sampled images. Then pre-trained LDMs are integrated to provide conditional priors in both latent and image domains. A novel Latent Guided Attention (LGA) is proposed for efficient fusion in multi-level latent domains. Simultaneously, to effectively utilize a prior in both the k-space and image domain, under-sampled images are fused with generated full-sampled images by the Dual-domain Fusion Branch (DFB) for self-adaption guidance. Lastly, to further enhance the data consistency, we propose a k-space regularization strategy based on the non-auto-calibration signal (NACS) set. Extensive experiments on two public MRI datasets fully demonstrate the effectiveness of the proposed methodology. The code is available at \url{https://github.com/Zolento/MDPG}.

\end{abstract}
\section{Introduction}
Magnetic Resonance Imaging (MRI) plays a crucial role in medical diagnostics owing to its non-invasive nature. However, its relatively time-consuming imaging process can lead to increased discomfort for the patients and pose a risk of diminished image quality. Compressed Sensing MRI (CS-MRI) reconstruction aims to recover original image signals from the under-sampled k-space acquisition, accelerating the imaging process while maintaining acceptable quality.

In recent years, generative models have been proven to outperform traditional numerical optimization algorithms in CS-MRI tasks. Recent generative adversarial networks (GANs) have been widely used in MRI reconstruction \cite{dagan, dlgan}. Despite the great success of GANs, unstable adversarial training resulting in gradient vanishing and mode collapse is often criticized. Thus, GAN may not be able to reconstruct accurate anatomical structures \cite{bau2019seeing}.

Diffusion models (DMs) use a parameterized Markov chain to replace the unstable adversarial training and have been widely applied to various image inverse problems with state-of-the-art (SOTA) performance \cite{ddpm, ddrm, wang2022zero}. One step further, latent diffusion models (LDMs) \cite{ldm} achieve the balance between generative quality and efficiency by transferring the diffusion process in the latent domain under a pre-trained variational autoencoder (VAE). With this feature, LDMs have become one of the most practical schemes for image generation.

Previous work has demonstrated the great potential of DMs in MRI reconstruction tasks \cite{zhao2024diffgan, bian2024diffusionqmri, xie2022measurement, prompting}. However, due to the random nature of DMs and the complicated anatomical structures in medical images, data consistency of DMs becomes an important factor in determining the reconstruction quality \cite{dps}. Though LDM as a novel improvement scheme can increase the efficiency of the original DM, naive data consistency in the k-space of generated images could add global artifacts. Moreover, operators in LDMs are highly non-convex, making the data consistency in the latent domain even more complicated \cite{rout2024solving, songsolving}.

Related work has shown that the LDM's compact prior knowledge of high dimensions helps networks learn the distribution of the original data more effectively. For instance, Li et al.\cite{li2024rethinking} utilizes a prior extraction module to compress target images into a highly compact latent domain where a conditioned denoising process is applied during inference; Luo et al.\cite{luo2024lare} applies latent reconstruction error guidance in the latent domain through single-step reconstruction to improve the efficiency of deep-generated image detection; Xie et al.\cite{xie2024learning} leverages an LDM learned from high-quality face images to provide features containing prior information for blind face restoration.

Inspired by the above works, we attempt to introduce prior guidance in MRI reconstruction tasks to improve reconstruction quality. Considering the compact nature of the latent domain and the proximity of the image domain to the ground-truth distribution, we focus on explicitly aligning the generated prior guidance with measurements using auxiliary modules. Our main contributions include:
\begin{itemize}
\item[--] We propose Multi-domain Diffusion Prior Guidance (MDPG), a two-stage approach integrated with prior knowledge in both latent and image domains provided by pre-trained LDMs. A novel Visual-Mamba-based backbone is utilized to process the measurements, i.e., the k-space of under-sampled images.
\item[--] We design the novel Latent Guided Attention (LGA) and Dual-domain Fusion Branch (DFB) modules to fully use prior knowledge guidance from different domains.
\item[--] We introduce a k-space regularization based on the non-auto-calibration signal (NACS) set representing the main detailed parts of images to further enhance the data consistency.
\end{itemize}
\begin{figure}[t]
    \includegraphics[width=1\textwidth]{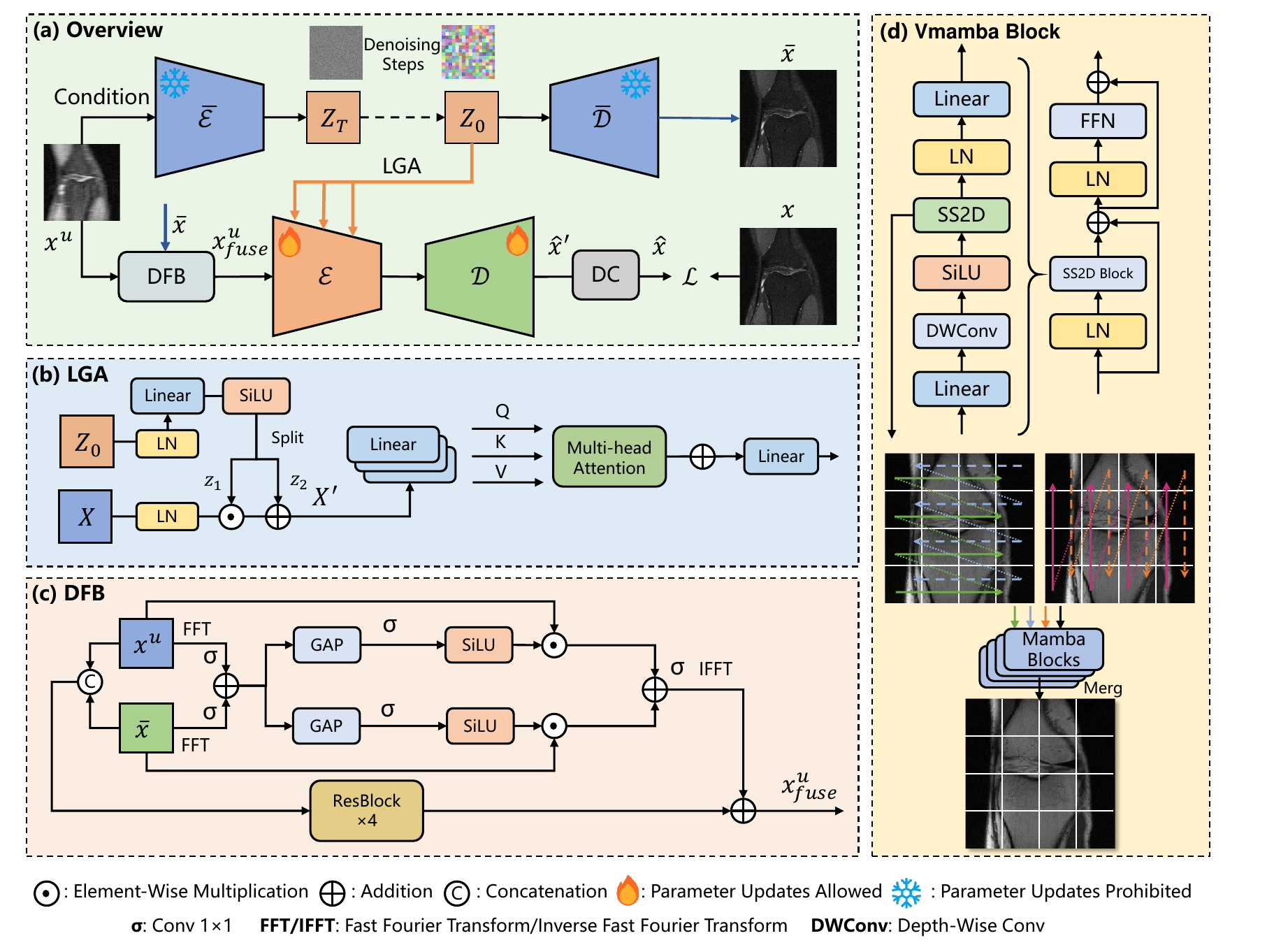}
    \caption{Overview of the proposed methodology. (a): The overall architecture of MDPG, where $\mathcal{E}$ and $\mathcal{D}$ represent the VMamba\cite{vmamba} encoder and convolutional decoder of the UNet-like backbone. DC refers to the data consistency layer. (b): The basic workflow diagram for LGA. (c): The structure of DFB, where the under-sampled image $x^{u}$ and synthesized full-sampled image $\overline{x}$ are fused and input into the backbone. (d): The structure and principles of VMamba blocks used in $\mathcal{E}$. SS2D refers to the 2D-Selective-Scan mechanism\cite{vmamba}.} 
    \label{fig:network}
\end{figure}
\section{Method}
We first consider the following forward sampling model for CS-MRI:
$$y=M\mathcal{FS}x+n, n \sim N(0, \sigma^2_y I),$$
where $x\in \mathbb{C}^{N}$ is the original full-sampled image, $y\in \mathbb{C}^{M}$ is the k-sapce measurement, and $n$ is the measurement noise. Here, $M$ is the under-sampling mask, $\mathcal{F}$ presents the Fourier Transform, and $\mathcal{S}$ denotes the coil sensitivity map. CS-MRI aims to reconstruct an estimated high-quality $\hat{x}$ from $y$:
\begin{equation}
\hat{x}=\arg\min_x\underbrace{||y-M\mathcal{FS}x||}_{\text{data consistency}}+\underbrace{\lambda\cdot\mathcal{R}(x)}_{\text{regularization}}.
\end{equation}
\subsection{Two-stage Strategy}
As illustrated in Fig.\ref{fig:network}, MDPG contains a Visual-Mamba-based backbone and a pre-trained LDM. We first train an LDM conditioned on the under-sampled image $x^{u}=\mathcal{F}^{-1}y$, then the backbone is trained under the guidance of LDM.
\subsubsection{Stage \uppercase\expandafter{\romannumeral1}}
 Given the under-sampled image $x^{u}$, the LDM encoder $\overline{\mathcal{E}}$ outputs a compact latent condition. We have the following conditional probability distribution\cite{ldm}:
\begin{equation}
    q\left(Z_{T}\mid Z\right)=\mathcal{N}\left(Z_{T};\sqrt{\overline{\alpha}_{T}}Z,\left(1-\overline{\alpha}_{T}\right)I\right),
\end{equation}
where $Z_{t}$ is the latent representation of $t$-th step, $T$ is the total number of diffusion steps and $\overline{\alpha}_{t}=\prod_{i=1}^{t}\alpha_{i}$, $ \alpha_{t}=1-\beta_{t}$ are the hyper-parameters that control the noise variance. The denoising process can be expressed as:
\begin{align}
    &q\left( Z_{t-1} \mid Z_{t}, Z_0 \right) = \mathcal{N}\left( Z_{t-1}; \mu_{t}\left(Z_{t}, Z_0\right), \frac{1 - \overline{\alpha}_{t-1}}{1 - \overline{\alpha}_{t}} \beta_{t} I \right),\\
    &\mu_{t}(Z_{t}, Z_0) = \frac{1}{\sqrt{\overline{\alpha}_{t}}} (Z_{t} - \frac{\beta_{t}}{\sqrt{1 - \overline{\alpha}_{t}}} \epsilon). 
\end{align}
Finally, the predicted noise $\epsilon_\theta (Z_{t}, x^{u}, t))$ provided by a time-conditional UNet\cite{unet} is used to predict $Z_{t-1}$ to $Z_{0}$ following the denoising process below: 
\begin{equation}
    Z_{t-1} = \frac{1}{\sqrt{\alpha_{t}}} (Z_{t} - \frac{1 - \alpha_{t}}{\sqrt{1 - \overline{\alpha}_{t}}} \epsilon_\theta (Z_{t}, x^{u}, t)) + \sqrt{1 - \alpha_{t}} \epsilon,
    \epsilon \sim \mathcal{N}(0, I).
\end{equation}
We follow the original DM to train the noise-predicting UNet $\epsilon_{\theta}(\circ,t)$, which optimizes the following target:
\begin{equation}
    \mathcal{L}_{\text{LDM}}:=\mathbb{E}_{\overline{\mathcal{E}}(x),\epsilon\sim\mathcal{N}(0,1),t}\left[\|\epsilon-\epsilon_\theta(Z_{t}, x^{u}, t)\|_2^2\right].
\end{equation}
\subsubsection{Stage \uppercase\expandafter{\romannumeral2}}
Once the LDM is trained, DDIM sampler\cite{ddim} and LDM decoder $\overline{\mathcal{D}}$ are used to generate $Z_{0}$ and synthesized full-sampled image $\overline{x}$. Given that down-sampling can create global artifacts in the image domain, a larger receptive field aids in feature extraction and fusion. Leveraging the innovative selective scan mechanism, the recently introduced VMamba model can extract global features from images with linear complexity. Consequently, we utilize VMamba for efficient processing of $Z_{0}$ and $\overline{x}$ within the encoder. Conversely, we opt for a straightforward convolutional design to restore fine-grained details while minimizing computational overhead, thereby providing sufficient precision in the decoder. 

As Fig.\ref{fig:network} demonstrates, $Z_{0}$ and $\overline{x}$ are integrated with the VMamba encoder $\mathcal{E}$ by the LGA and DFB modules, respectively. The architecture of the backbone follows the UNet\cite{unet}, where the convolution layers in the original encoder are replaced with VMamba layers which follow the design in Fig.\ref{fig:network}(d). The final loss function is described in Eq.\ref{loss}.
\subsection{Prior Knowledge Guidance}
The compact prior $Z_{0}$ and corresponding decoded synthesized full-sampled image $\overline{x}$ are given during the training process of the backbone. For $Z_{0}$, LGA is proposed to inject prior knowledge into the backbone encoder $\mathcal{E}$. Let $W$ denote linear projection layers, LN refer to the layer-norm operation, and $X$ represent the feature in the corresponding layer of $\mathcal{E}$, LGA can be expressed as:
\begin{equation}
\begin{aligned}
    &\mathrm{LGA(Q,K,V)}=\text{Softmax}(\frac{QK^T}{\sqrt{d_k}})V,\\
    &\text{where }Q,K,V=X'W^{Q},X'W^{K},X'W^{V}, X'=LN(X)\cdot z_{1} + z_{2},\\
    &z_{1}, z_{2} = \text{Split}(\text{SiLU}(\text{LN}(Z_{0})W^{Z})).
\end{aligned}
\end{equation}
The k-space as the measurement space of MRI signals contains a wealth of original frequency information, which helps to enhance data consistency. Thus, we design DFB to extract the rich details from both the image domain and frequency domain, as shown in Fig.\ref{fig:network}. Considering the input image's under-sampled characteristic in the k-space, a dynamic weighting strategy is chosen for complementarity. Meanwhile, we use a residual branch to extract information from the concatenated image features. The detailed process is as follows:
\begin{align}
    &x^{u}_{fuse}=\sigma_{iout}(\mathcal{F}^{-1}[f(\sigma_{x}(x^{u}_{k}), \sigma_{y}(\overline{x}_{k}))] + g(x^{u}, \overline{x})),\\
    &f(x, y)=\sigma_{kout}[x\cdot \text{SiLU}(t_{1}(x, y))+y\cdot \text{SiLU}(t_{2}(x,y))],\\
    &t_i(x, y)=\sigma_i (\phi(x+y)),\\
    &g(x, y)=\text{ResBlocks}(\text{concat}(x, y)),
\end{align}
where $\sigma(\cdot)$ represents the 1$\times$1 convolution and $\phi(\cdot)$ denotes the global average pooling (GAP) operation, respectively. The subscript $k$ indicates that the feature is transformed to the frequency domain, i.e., k-space. The output features fused by the DFB are fed again into the encoder $\mathcal{E}$ as input to the backbone network.
\subsection{NACS Set based K-space Regularization}
Degradation modes in under-sampled images are closely related to the under-sampling mask. The data consistency layer is often achieved by a learnable parameter $\nu$ related to noise level $n$\cite{dccnn}:
\begin{equation}
    \left.\hat{x}_{k}(i_k,j_k)=\left\{\begin{array}{ll}\hat{x}^{'}_{k}(i_k,j_k),&(i_k,j_k)\notin\Omega\\\frac{\hat{x}^{'}_{k}(i_k,j_k)+\nu x^{u}_{k}(i_k,j_k)}{1+\nu},&(i_k,j_k)\in\Omega\end{array}\right.\right.
\end{equation}
$\Omega$ represents the under-sampled k-space set and $\hat{x}^{'}_{k}$ denotes k-space of the original output of the backbone. This approach results in the optimization being dominated by the main low-frequency signal represented by the ACS. We recommend optimizing k-space separately according to the NACS set which represents most of the high-frequency, i.e., detailed parts of images. The combined loss function is defined as:
\begin{equation}
\label{loss}
    \mathcal{L}=\frac{\|\hat{x}-x\|_2}{\|x\|_2} + \lambda_{1}\frac{\|\hat{x}^{\Gamma}_{k}-x^{\Gamma}_k\|_2}{\|x^{\Gamma}_k\|_2}
\end{equation}
The first term denotes minimizing the loss in the image domain. The second term regularizes the k-space according to the sampling pattern. Here $\Gamma$ denotes the NACS set. $\lambda_{1}$ is a fixed hyperparameter.
\section{Experiments and Results}
\begin{table*}[t]
    \centering
    \setlength{\tabcolsep}{6pt}
    \caption{Reconstruction results of FastMRI dataset. Best results are bolded while second best are underlined. }
    \begin{tabular}{l|c|c|c|c|c|c}
        \hline
         \multirow{2}{*}{Method} &\multicolumn{3}{c|}{8$\times$ 0.04c}&\multicolumn{3}{c}{4$\times$ 0.08c} \\
         \cline{2-7}
         &PSNR$\uparrow$&SSIM$\uparrow$&NMSE$\downarrow$&PSNR$\uparrow$&SSIM$\uparrow$&NMSE$\downarrow$\\
         \hline
         ZF (Zero-Filling)&26.82  &0.619  &0.0893  &29.47  &0.729  &0.0531 \\
         ISTA-Net \cite{ista}&\secondbest{29.75}  &0.724  &\secondbest{0.0502}  &31.54  &0.759  &0.0410 \\
         DCCNN \cite{dccnn}&29.25  &0.710  &0.0546  &\secondbest{31.97}  &0.787  &0.0346 \\
         MD-Recon \cite{dccnn}&29.52  &0.717  &0.0521  &31.94  &\secondbest{0.789}  &\secondbest{0.0343} \\
         UNet \cite{unet}&29.51  &\secondbest{0.725}  &0.0546  &31.68  &0.787  &0.0382 \\
         \hline
         DAGAN \cite{dagan}&28.40  &0.684  &0.0635  &30.91  &0.766  &0.0407 \\
         DDNM \cite{wang2022zero}&27.89  &0.665  &0.0833  &31.35  &0.762  &0.0418 \\
         LDM-16 \cite{ldm}&25.99  &0.536  &0.113  &26.98  &0.578  &0.0945 \\
         MC-DDPM \cite{xie2022measurement}&27.35  &0.614  &0.0900  &29.16  &0.683  &0.0658 \\
         \hline
         \textbf{MDPG (ours)}&\best{30.43}  &\best{0.740}  &\best{0.0451}  &\best{32.14}  &\best{0.794}  &\best{0.0334} \\
         \hline
    \end{tabular}
    \label{fastmri}
\end{table*}
\begin{table*}[t]
    \centering
    \setlength{\tabcolsep}{6pt}
    \caption{Reconstruction results of IXI T1 dataset. The middle 10 slices of each volume were used to calculate the metrics. Best results are bolded while second best are underlined.}
    \begin{tabular}{l|c|c|c|c|c|c}
        \hline
         \multirow{2}{*}{Method} &\multicolumn{3}{c|}{8$\times$ 0.04c}&\multicolumn{3}{c}{4$\times$ 0.08c} \\
         \cline{2-7}
         &PSNR$\uparrow$&SSIM$\uparrow$&NMSE$\downarrow$&PSNR$\uparrow$&SSIM$\uparrow$&NMSE$\downarrow$\\
         \hline
         ZF (Zero-Filling)&21.75  &0.529  &0.103  &24.45  &0.668  &0.0554 \\
         ISTA-Net \cite{ista}&\secondbest{26.97}  &\secondbest{0.829}  &\secondbest{0.0313}  &\best{32.37}  &\best{0.942}  &\best{0.00915} \\
         MD-Recon \cite{mdrecon}&26.90  &0.813  &0.0317  &\secondbest{32.01}  &\secondbest{0.932} &\secondbest{0.00988} \\
         UNet \cite{unet}&26.47  &0.796  &0.0353  &30.06  &0.891  &0.0155 \\
         \hline
         DAGAN \cite{dagan}&24.30  &0.735  &0.0573  &25.86  &0.787  &0.0400 \\
         DDNM \cite{wang2022zero}&24.11  &0.747  &0.0607  &29.10  &0.870  &0.0191 \\
         LDM-16 \cite{ldm}&20.99  &0.600  &0.122  &22.56  &0.679  &0.0854 \\
         MC-DDPM \cite{xie2022measurement}&23.97  &0.672  &0.0624  &28.21  &0.800  &0.0239 \\
         \hline
         \textbf{MDPG} (ours)&\best{27.50}  &\best{0.833}  &\best{0.0278}  &31.00  &0.914 &0.0124 \\
         \hline
    \end{tabular}
    \label{IXI}
\end{table*}
\subsection{Datasets and Baselines}
We evaluate our method on two MRI datasets: FastMRI\footnote{\url{https://fastmri.med.nyu.edu/}} single-coil knee dataset and IXI\footnote{\url{https://brain-development.org/ixi-dataset/}} T1 dataset. The FastMRI dataset consists of a training set with 973 volumes and a validation set with 199 volumes. The IXI T1 dataset is divided into a training set with 462 volumes and a validation set with 115 volumes. For preprocessing, each of the slices in the FastMRI dataset is cropped into 320$\times$320 size, while those in the IXI T1 dataset are cropped into 256$\times$256. Under-sampled k-space was generated by applying 4$\times$ and 8$\times$ Cartesian masks on full-sampled k-space using official codes of FastMRI.

We compare our method with different types of MRI reconstruction baselines. For end-to-end methods, we choose ISTA-Net\cite{ista}, UNet\cite{unet}, and MD-Recon-Net\cite{mdrecon}. For generative models, we choose DAGAN\cite{dagan}, LDM\cite{ldm}, DDNM\cite{wang2022zero}, and MC-DDPM\cite{xie2022measurement}. Average metrics are computed on every volume of the validation set, including Peak Signal-to-Noise Ratio (PSNR), Structural Similarity Index Measure (SSIM), and Normalized Mean Square Error (NMSE) values.
\begin{figure}[t]
    \includegraphics[width=1\textwidth]{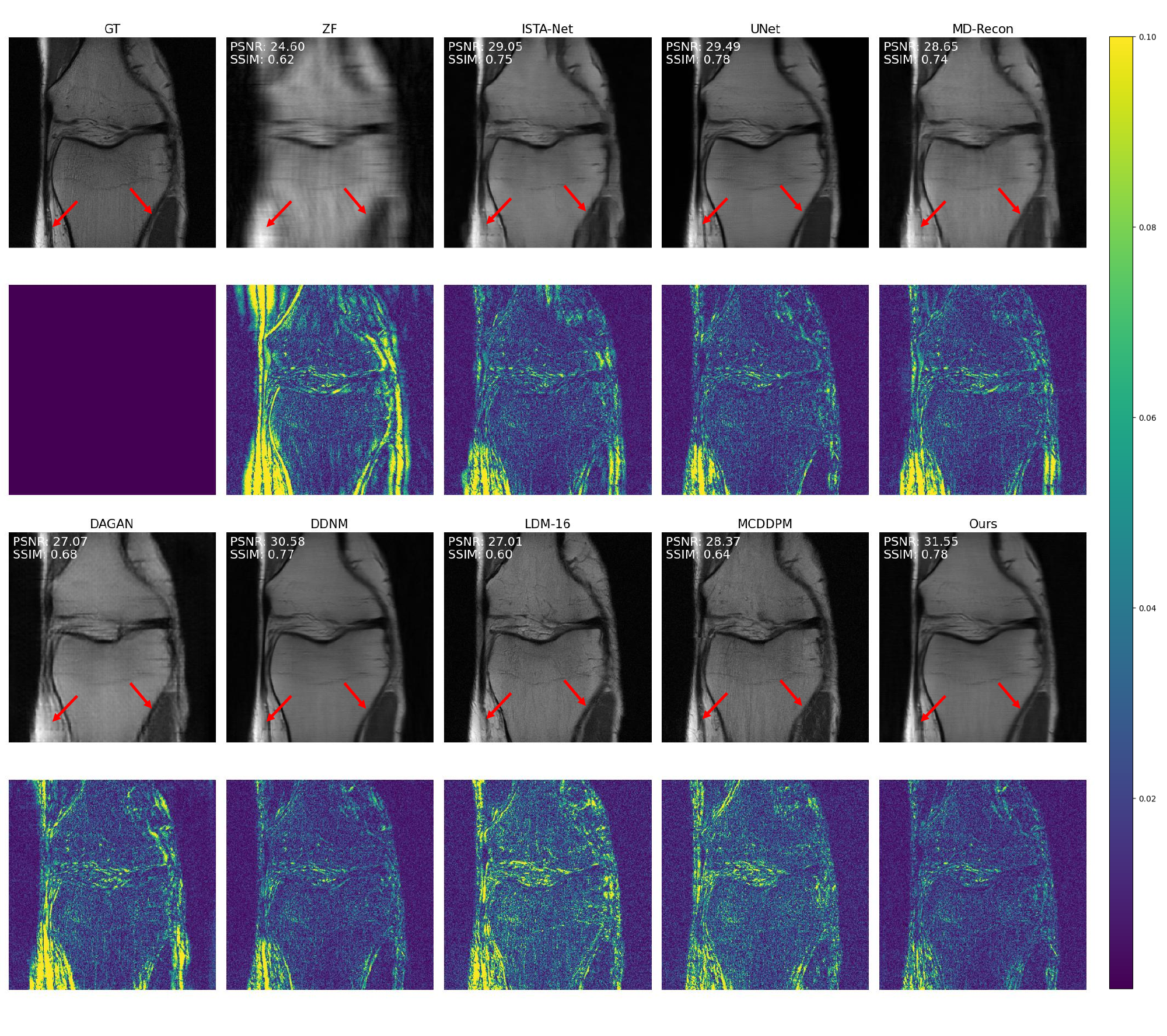}    \caption{Visualization of reconstruction results and corresponding error maps of FastMRI dataset of 8$\times$ acceleration factor. Red arrows indicate highly degraded key anatomical structures that are desired to be as close as possible to the ground truth (GT).}
    \label{fig:fastmri}
\end{figure}
\subsection{Implementation Details}
We trained MDPG's backbone with learning rate = 4$\times$10$^{-3}$, batch size = 8, and an AdamW optimizer for 100 epochs. LDM was trained with $T=1000$, 16$\times$ down-sampled $Z$, and a linear noise schedule. The sampling step of DDIM is set to 20 in both the training and inference processes. $\lambda_{1}$ is set to 0.1. All experiments are done with an NVIDIA GeForce RTX 4090 GPU under the Pytorch framework.
\subsection{Results}
\begin{table*}[t]
    \centering
    \setlength{\tabcolsep}{8pt}
    \caption{Ablation results of components and training configurations in MDPG under 8$\times$ acceleration of the FastMRI dataset. A: DFB; B: LGA; C: using Sigmoid instead of SiLU in DFB; D: learnable $\nu$; E: placing DFB before $\mathcal{E}$ instead of after $\mathcal{D}$. F: NACS set regularization. The first row represents the results of the backbone without prior guidance.}
    \begin{tabular}{c|c|c|c|c|c|c|c|c}
        \hline
         \textbf{A} &\textbf{B} &\textbf{C} &\textbf{D} &\textbf{E} &\textbf{F} &PSNR$\uparrow$&SSIM$\uparrow$&NMSE$\downarrow$\\
         \hline
         N/A &N/A  &N/A  &\Checkmark  &N/A  &\XSolidBrush &29.97  &0.731  &0.04870\\
         \Checkmark &\Checkmark  &\XSolidBrush  &\Checkmark  &\XSolidBrush  &\XSolidBrush &29.56  &0.679 &0.05776 \\
         \Checkmark &\Checkmark   &\Checkmark  &\Checkmark  &\XSolidBrush  &\XSolidBrush &30.10  &0.730 &0.04812 \\
         \Checkmark &\Checkmark   &\Checkmark  &\XSolidBrush  &\Checkmark  &\XSolidBrush &30.23  &0.735 &0.04700 \\
         \Checkmark &\Checkmark   &\XSolidBrush  &\Checkmark  &\Checkmark  &\XSolidBrush &30.12  &0.733 &0.04745 \\
         \Checkmark &\XSolidBrush   &\Checkmark  &\Checkmark  &\Checkmark  &\XSolidBrush &30.20  &0.736 &0.04676 \\
         \XSolidBrush &\Checkmark  &\Checkmark  &\Checkmark  &\Checkmark  &\XSolidBrush &30.24  &0.738 &0.04627 \\
          \Checkmark &\Checkmark  &\Checkmark  &\Checkmark  &\Checkmark  &\XSolidBrush &\secondbest{30.33}  &\secondbest{0.739} &\secondbest{0.04614} \\
         \Checkmark &\Checkmark  &\Checkmark  &\Checkmark  &\Checkmark  &\Checkmark &\best{30.43}  &\best{0.740} &\best{0.04514} \\
         \hline
    \end{tabular}
    \label{ablation}
\end{table*}
As Table \ref{fastmri} and \ref{IXI} demonstrate, our method achieves the best performance in three scenarios. In the complex-valued FastMRI dataset, MDPG has universal advantages in all metrics. Fig.\ref{fig:fastmri} demonstrates how our method outperforms other baselines. MDPG restores two highly degraded anatomical structures more accurately than end-to-end baselines without introducing excessive false details like generative models. In other regions, MDPG also demonstrates excellent performance without exhibiting significant errors. For the real-valued IXI T1 dataset, MDPG is more advantageous under 8$\times$ acceleration. Though it does not yield optimal results at 4$\times$ factors, the performance was still better than all generative baselines. The performance degradation is most likely attributed to the low quality of the generated prior, as evidenced by the low metrics for LDM under 4$\times$ acceleration. 
\subsection{Ablation Study}
We conducted ablation experiments to verify the effectiveness of each key component and configuration in MDPG. As Table \ref{ablation} shows, priors in the latent, image, and frequency domains improve performance over the backbone. Experiments also demonstrate performance improvement by the NACS set regularization.
\section{Conclusion}
This paper introduces MDPG, a novel approach incorporating a latent diffusion prior, a multi-domain fusion strategy, and k-space regularization based on the NACS set. The experiments demonstrate that our method outperforms other baseline approaches. In conclusion, MDPG illustrates that the utilization of multi-domain diffusion prior knowledge could effectively enhance the performance of MRI reconstruction tasks. 
\begin{credits}
\subsubsection{\discintname}
The authors have no competing interests to declare that are relevant to the content of this article.
\end{credits}
\bibliographystyle{splncs04}
\bibliography{Paper-2241}
\end{document}